\documentclass[11pt,twoside]{article}

\usepackage{asp2004}
\usepackage{epsf}
\usepackage{psfig}
\usepackage{lscape}
\usepackage{graphicx}

\def\mum{${\rm \mu m}$}
\newcommand{\HII}{\ion{H}{ii}~}

\begin{document}
\title{The 15-21 $\mu$m PAH plateau} 

\author{E. Peeters$^1$, A.L. Mattioda$^1$, F. Kemper$^2$,
D.M. Hudgins$^1$, L.J. Allamandola$^1$}

\affil{$^1$NASA Ames Research Center, MS245-6, Moffett Field,
CA94035\\ $^2$University of Virginia, PO Box 3818, Charlottesville, VA
22903}

\begin{abstract} 
We present 14--21 \mum\, emission spectra of star forming regions and
compare them with the PAH spectra from the Ames PAH database. We show
that while the emission in this region can be quite variable, the bulk
of these variations can be accommodated by variations in the IS PAH
population.
\end{abstract}

\section{Introduction}
The mid-IR spectra of many sources are dominated by emission bands at
3.3, 6.2, 7.7, 8.6, and 11.2 \mum, generally attributed to Polycyclic
Aromatic Hydrocarbons (PAHs). It is now firmly established that the
characteristics of the PAH bands vary from source to source and
spatially within sources revealing much about the carrier and the
local physical conditions \citep[for a review see][]{Peeters:review:04}.

\citet[][ATB]{Allamandola:rev:89} pointed out that objects showing PAH
bands are expected to show emission at wavelengths longer than
$\sim$15 \mum\, arising from their CCC out-of-plane bending
vibrations. This was confirmed by ISO observations. New features were
reported at 16.4 (\citeauthor{Moutou:16.4:00}
\citeyear{Moutou:16.4:00}; \citeauthor{VanKerckhoven:plat:00}
\citeyear{VanKerckhoven:plat:00}, VHP) and 17.4 \mum\,
\citep{VanKerckhoven:phd:02} while VHP present evidence for a
variable 15--21 \mum\, plateau.  These authors interpreted
the plateau as ``a collection of blended emission features''. Using Spitzer
further details about this variable plateau were revealed (Spitzer Special
Issue, ApJS).

\section{Results}

\begin{figure}[!t]
\plottwo{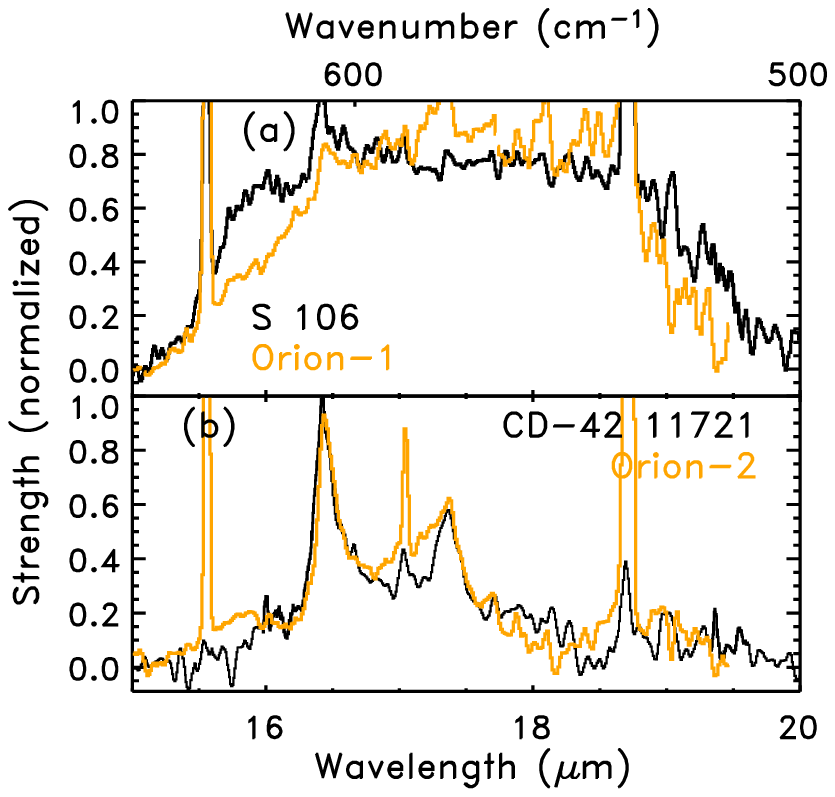}{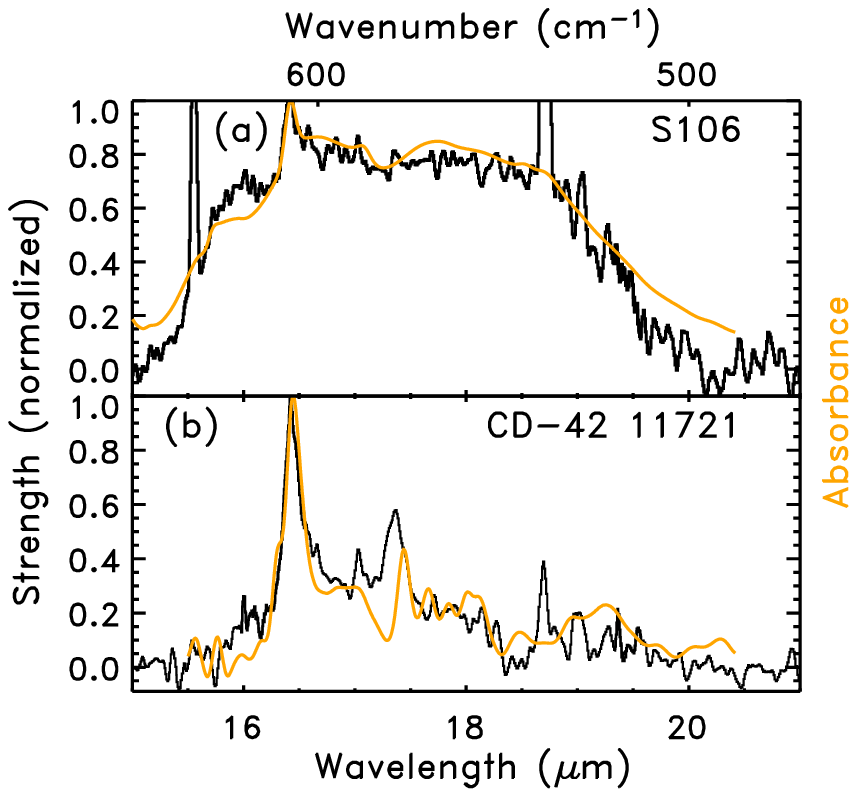}
\caption{The 14 to 20 \mum\, emission spectra of S~106 and CD~-42
11721 (ISO-SWS; VHP). These are compared with 2 positions within Orion
(Spitzer-IRS; Peeters et al. in prep., left panel) and
with PAH laboratory spectra produced by co-adding the spectra of
individual, neutral PAHs and applying a 10\,$cm^{-1}$ redshift to
account for the emission process (PMH, right panel).}
\label{obs}
\end{figure}

VHP showed that there are significant variations in the 15-21 \mum\, PAH emission. Fig.~\ref{obs} shows their two most extreme cases : while the
emission in CD~-42~11721 is dominated by bands at 16.4 and 17.4 \mum,
S~106 shows a broad, nearly flat-topped plateau between 15 to 21
\mum\, with a secondary 16.4 \mum\, band (VHP). In addition, there is
much observational evidence of discrete emission features at 15.8,
16.4, 17.4 and 19.0 \mum\, \citep[see e.g. VHP;][PMH, and references therein]{Peeters:plat:04}
which can vary spatially as observed for NGC~7023
\citep{Werner:04}. Emission in this region is thus comprised of at
least two classes of independent components: distinct, narrow features
and broad variable plateaus. In addition, \HII regions seem to show
the broad variable plateau while YSO's and reflection nebulae show a
complex of narrow features.

Fig.~\ref{obs} shows Spitzer-IRS observations of 2 positions in the
Orion Bar. Clearly, within the Orion Bar, both the broad plateau and
the complex of features are observed. The broad plateau is located
closer to the ionizing stars compared to the complex of
features. Combined with the dependence of the emission on object type
(see above), this suggests that either the carrier of the broad
plateau is more stable against UV photons than the carrier of the
complex of features or the ionization state of the carrier of both
features is different.


Most spectra in the Ames PAH database have bands in the 15--21 \mum\,
range arising from out-of-plane skeletal vibrations.  Fig.\ref{obs}
compares the astronomical spectra with different PAH mixtures. Note
that these mixtures are not unique.  The principle purpose here is to
demonstrate that band blending from different PAHs in a mixture can
readily produce these features (ATB, Moutou et al. 2000, VHP,
PMH). The discrete narrow bands are likely due to gas-phase PAH
molecules while emission from larger PAHs, PAH clusters etc. likely
blend and primarily contribute to the broad, somewhat structureless
underlying plateau.

{\it Details on the laboratory comparison and its implications
can be found in PMH. The Spitzer data of Orion will be presented in
Peeters et al. (in prep.).}

\acknowledgements This work was supported by NASA's Long
Term Space Astrophysics Program, the Spitzer Fellowship Program and
the National Research Council.


\begin{thebibliography}{}

\bibitem[{Allamandola} et~al.(1989)]{Allamandola:rev:89}
L.~J. {Allamandola}, A.~G.~G.~M. {Tielens}, J.~R. {Barker} 1989,
\apjs, 71, 733 (ATB)

\bibitem[{Moutou} et~al.(2000)]{Moutou:16.4:00} C.~{Moutou},
  L.~{Verstraete}, A.~{L{\' e}ger}, K.~{Sellgren}, W.~{Schmidt}
  2000, \aap, 354, L17

\bibitem[{Peeters} et~al.(2004{\natexlab{a}})]{Peeters:review:04}
E.~{Peeters}, L.~J. {Allamandola}, D.~M. Hudgins et al. 2004{\natexlab{a}}, In A.~N. {Witt}, editor, {\em Astrophysics of Dust}, Astronomical
  Society of the Pacific, p141

\bibitem[{Peeters} et~al.(2004{\natexlab{b}})]{Peeters:plat:04}
E.~{Peeters}, A.L.~{Mattioda}, D.M.~{Hudgins}, L.~{Allamandola} 2004b,
\apjl, 617, L65, (PMH)

\bibitem[{Van Kerckhoven}(2002)]{VanKerckhoven:phd:02}
C.~{Van Kerckhoven} 2002, PhD thesis, Katholieke Universiteit Leuven (Belgium)

\bibitem[{Van Kerckhoven} et~al.(2000)]{VanKerckhoven:plat:00}
C.~{Van Kerckhoven}, S.~{Hony}, E.~{Peeters} et al. 2000, \aap, 357, 1013 (VHP)

\bibitem[{Werner} et~al.(2004)]{Werner:04}
M.~W. {Werner}, K.~I. {Uchida}, K.~{Sellgren} et al. 2004. \apjs, 154, 309

\end{thebibliography}
\end{document}